\documentclass[aps12,pre,twocolumn,preprintnumbers,amsmath,amssymb,floatfix]{revtex4}
\usepackage[dvips]{graphicx}

\DeclareGraphicsExtensions{.eps,.ps}

\def\s#1{_{\rm #1} }

\def\trans{\textit{trans}\,}
\def\cis{\textit{cis}\,}
\def\half{{\textstyle \frac{1}{2}}}

\def\cI{ {\cal I } }
\def\cR{ {\cal R } }
\def\cC{ {\cal C } }
\def\bea{\begin{eqnarray}}
\def\eea{\end{eqnarray}}
\def \be{\begin{equation}}
\def \ee{\end{equation}}
\def\nl{\hfil\break}
\begin{document}

\title{Bleaching and stimulated recovery of dyes and of photo-cantilevers}
\author{D. Corbett and M. Warner}\email{mw141@cam.ac.uk}
\affiliation{Cavendish Laboratory, Madingley Road, Cambridge, CB3 0HE, U.K.}

\date{\today}
\begin{abstract} We examine how intense optical beams can penetrate deeply into highly absorbing media by a non-linear, photo-bleaching process.  The role of stimulated recovery to the dye ground state can be important and is delineated.  This analysis of non-linear absorption processes is applicable in general to situations where chromophores are irradiated, for instance in biology.  We examine the implications for the bending of cantilevers made of heavily dye-loaded nematic photo-solids, that is nematic glasses and elastomers that have large mechanical reactions to light.  In particular we describe the bending of cantilevers sufficiently absorbing that they would not bend if Beer's Law were applicable.  We quantify the role of optically-generated heat in determining the mechanical response and conclude that in general it is minor in importance compared with optical effects.

 \end{abstract}
 \maketitle

\section{Introduction}
\begin{figure}[t]
\centering\resizebox{0.3\textwidth}{!}{\includegraphics{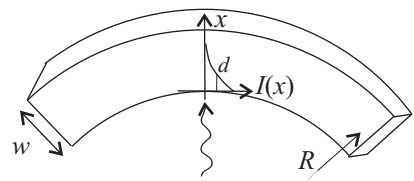}} \caption[]{Radiation penetration  with linear absorption
length $d$ giving light-induced bend.} \label{fig:slender}
\end{figure}

Nematic solids are principally elastomeric \cite{warnerbook:03} or glassy \cite{Mol:05}.  In each case they are composed of nematic polymers, but either loosely or tightly crosslinked.  Their moduli are accordingly low or high and they are capable of either huge or modest extensions.  They form unique solids in that, when their order is changed, they can change their dimensions considerably and these mechanical responses can be steered in many ways.  The order can either be redirected (director rotation under the influence of electric, mechanical or optical fields) or reduced.  In this paper, for simplicity we concentrate on  changing the magnitude of order, achievable  by heating or cooling, or by the absorption of light into photo-active rods when they are a component of the nematic solid.  The natural elongation of the oriented polymers making up the networks is modified by order change, with the result that macroscopic mechanical changes are induced. Reversible thermal strains of hundreds of per cent have been observed in elastomers \cite{Wermter,Loft}.  Smaller changes arise in glasses, where subtle arrangements of director have been employed to achieve more complex mechanical responses \cite{Mol:05}.  Since light and heat have analogous effects, one expects and indeed finds that optical absorption leads to analogous mechanical strains, both in photo-elastomers \cite{Finkphoto,Cviklinski:03} and in photo-glasses \cite{vanOosten:07b}.  Such a mechanism of mechanical change offers the possibility of \textbf{m}icro-\textbf{o}pto-\textbf{m}echanical
\textbf{s}ystems (MOMS) where elements can be optically-induced to
bend as elastomeric photo-swimmers \cite{Palffy-Muhoray:04} or as
glassy cantilevers \cite{Harris:05,Tabiryan:05,vanOosten:07b}.  The bend arises because the contraction is differential with depth since the optical beam is absorbed and is hence weaker with depth.  How the beam intensity varies with depth because of non-linear absorption processes, and how this variation is translated into bending even in situations where absorption is so high that penetration would be negligible in the linear case, are the two questions that concern this paper.  A schematic, Fig.~\ref{fig:slender}, of the penetration and of the differential contraction leading to bend establishes the coordinate ($x$) of penetration, the thickness ($w$), and the curvature ($1/R$) of the cantilever. 

Light disrupts nematic order when molecular rods which are also chromophores, that is
dye molecules, bend on absorbing a photon in their straight (\trans)
ground state and make an indirect transition to their bent (\cis) excited state.  The orientational order of tightly packed rods is thereby reduced by the \cis fraction.  We shall assume for simplicity that the mechanical contractile strain is proportional to the \cis volume fraction.  It is a vexed question as to whether heat released by these optically-induced transitions is the actual cause of the order change.  This question has been addressed by irradiating polydomain nematic photo-glasses with polarized light.  One finds \cite{Ikeda:03} that the mechanical contraction occurs along the light polarization direction, a response that is therefore simply tuneable by rotating the light polarization.  Domains closely aligned to the optical electric field absorb, and thus also contract, preferentially over domains that are misaligned \cite{Corbett:06}.  Moreover, since the domain size is small, heat diffusion is fast and domains neighboring an absorbing region quickly reach similar temperatures.  If temperature were the mechanism for contraction, it would thus under these conditions of fast diffusion not distinguish a preferred direction and there would be no overall effect (just as when one simply heats a polydomain nematic solid).  Evidently, optical effects dominate.  We discuss thermal effects at length in this paper and find their influence  small, even across a whole monodomain  cantilever, because of the diffusional rates that arise.

Since photons are absorbed by the dye components in the nematic, then light penetrating the face of a nematic
photo-cantilever will be attenuated and hence the contractions generated diminish with depth.  Curvature of
the cantilever results, Fig.~\ref{fig:slender}.
It is important in \textbf{m}icro-\textbf{o}pto-\textbf{m}echanical
\textbf{s}ystems (MOMS) where elements can be optically-induced to
bend as elastomeric photo-swimmers \cite{Palffy-Muhoray:04} or as
glassy cantilevers \cite{Harris:05,Tabiryan:05}.

Weak beams decay
exponentially with depth (Beer's law).  The conversion of straight to bent
(\textit{trans} $\rightarrow$ \textit{cis}) forms of the dye
molecules is also exponential in this limit. For a linear connection (valid for small strains)
between \textit{cis} population and contraction,  in the Beer limit maximum cantilever curvature is predicted
\cite{Mahadevan04} for $w/d \sim 2.63$, where $w$ is the thickness
of the cantilever and $d$ is the exponential decay length: if $w \gg
d$, only a thin skin of network contracts and its contractile
stresses are insufficient to make the unstrained part of the
cantilever below respond. Equally, if $w \ll d$, then there is
little variation of photo-strain through the thickness and the
cantilever may contract but not differentially with depth and thus
will  bend little.  The extent of bend, for a fixed $w/d$, was also
predicted to be linear with intensity.

Experimentally, cantilevers are commonly heavily dye-doped but still
show appreciable mechanical effects \cite{Kondo:06}.  The above
arguments suggest that this bending is unexpected: high dye
concentrations mean strong absorption and hence small $d$.  In this
thin-skin limit, $d \ll w$, bend is expected to disappear. Evidently
non-linear effects lead to deep penetration of light and thus lead
to bending. We shall explain these related effects by making two
assumptions for simplicity to illustrate the underlying principles.
We take the straight forward case of no photo-induced director
rotation, that is for nematic glasses  and constrained systems (for
instance where surface effects might be strong, where the director
seems to be immobile under elongations imposed at an angle and also
probably don't rotate during photo processes \cite{vOosten:07}.
However, photoisomerisation in glasses has additional subtleties; we
derive the non-linear penetration of intense beams into glass as
well. Secondly we assume that there is not appreciable reduction in
the magnitude of the nematic order parameter under illumination.
This assumption is reasonable in glasses, but in liquid and
elastomeric nematics, this assumption is too restrictive.  Indeed a
description of polydomain nematic elastomer response invokes both
order reduction and rotation \cite{Corbett:06}. We return elsewhere
to the role order reduction and rotation play in non-linear
absorption and mechanics.

An initial attack on this problem \cite{Corbett:07} invoked
photo-bleaching, that is depletion of the \trans form, letting light
through to greater depths than would be expected by Beer's law. Here
we derive these non-linear effects fully, paying attention to two
additional influences that are potentially important.  The
optically-stimulated back reaction \cis $\rightarrow$ \trans can
alter the non-linear processes considerably and is of importance
when the \cis absorption line is not widely separated from that of
the \trans species. Secondly, thermal effects could be considerable,
all the more so because of a possibly nearby nematic to isotropic
transition where giant thermo-mechanical effects are known to occur,
especially in elastomers.

 The
remainder of this paper is organized as follows.  In section
\ref{sec:attenuation} we derive equations which describe the
attenuation of light passing through a region with photo-active
chromophores.  Our analysis of non-linear absorption is thus
relevant to a wide range of situations where dye is irradiated, and
is not limited to mechanics, which is our ultimate aim here. Our
methods and results are similar to a penetrating analysis,
experimental and theoretical, of Statman and Janossy
\cite{Statman:03}.  In section \ref{sect:curvature} we show how
optical attenuation leads to cantilever bending and calculate the
radius of curvature as a function of the incident flux of light.  In
section \ref{sec:neutrals} we investigate the distribution of strain
throughout the bent sample, and in particular we demonstrate that
there can be several  planes within the cantilever on which the net
strain is zero. In section \ref{sec:temperature} we discuss the
possible effects of temperature change owing to absorption of
photons on the results presented thus far.  We find that, for
intense illumination, temperature distributions are symmetric about
the cantilever mid-plane and hence do not contribute to bend, only
contraction.  Finally in \ref{sec:conclusions} we present our
conclusions.

\section{Attenuation}\label{sec:attenuation}

We consider the situation of Fig.~\ref{fig:slender}(b) of a long,
slender cantilever of thickness $w$ illuminated by light with
incident flux $I\s{0}$.  The absolute number density of chromophores
is $\rho_{ph}$.  At a time $t$  after the onset of illumination and
depth $x$ within the cantilever, the fraction of these chromophores
in the straight \trans state is $n\s{t}(x,t)$ and the fraction in
the bent \cis state is $n\s{c}(x,t)=1-n\s{t}(x,t)$. The magnitude of
the Poynting flux at $x$
 and  $t$ is $I(x,t)$.  The dynamics of the
\textit{trans} fraction is determined by three processes, (i) an
optically stimulated \textit{trans} $\rightarrow$ \textit{cis}
reaction with rate $\Gamma_{1}I(x,t)n\s{t}(x,t)$, (ii) an optically
stimulated \textit{cis} $\rightarrow$ \textit{trans} back-reaction
with rate $\Gamma_{2}I(x,t)n\s{c}(x,t)$ and (iii) a spontaneous,
thermally activated, \textit{cis} $\rightarrow$ \textit{trans}
back-reaction with characteristic time $\tau$.  $\Gamma\s{t}$ and
$\Gamma\s{c}$ subsume absorption cross sections per chromophore and
the quantum efficiencies $\Phi_{tc}$ and $\Phi_{ct}$ of the
stimulated \textit{trans}-\textit{cis} reaction and
\textit{cis}-\textit{trans} back-reaction respectively, see
\cite{Statman:03} for the rate equations in full with such factors
explicitly given. We take the absorption cross sections  to be
independent of nematic order -- as discussed above, changing nematic
order is itself another source of non-linearity. Combining these
three rates we obtain for the rate of change of the \textit{trans}
fraction:
\begin{eqnarray}
\frac{\partial n\s{t}}{\partial t}=
-\Gamma\s{t}I(x,t)n\s{t}(x,t)+\left(\frac{1}{\tau}+\Gamma\s{c}I(x,t)\right)n\s{c}(x,t)\label{eq:dynamics}\end{eqnarray}
In this paper we confine ourselves to the steady-state, that is
$\frac{\partial n\s{t}}{\partial t}=0$.  Setting this condition in
eqn~(\ref{eq:dynamics}) and taking out a factor of $\tau$ gives the
steady state \textit{trans} and \textit{cis} populations:
\begin{equation}
\begin{array}{ccc}
n\s{t}(x)=\frac{1+\Gamma\s{c}\tau I}{1+(\Gamma\s{t}+\Gamma\s{c})\tau
I}&;&n\s{c}(x)=\frac{\Gamma\s{t}\tau I}{1+(\Gamma\s{t}+\Gamma\s{c})\tau
I}
\end{array}
\end{equation}
where $I$ is now $I(x)$ simply, and will be determined below. We can
identify two characteristic, material intensities,
$I\s{t}=1/(\Gamma\s{t}\tau)$ and $I\s{c}=1/(\Gamma\s{c}\tau)$.  It is
convenient to scale the flux by its incident value, thus
$\cI(x,t)=I(x,t)/I_{0}$. The reduced intensity is thus $\cI = 1$ at
the entry surface $x=0$, see Fig.~\ref{fig:slender}(b).  We also
define dimensionless quantities $\alpha=I_{0}/I\s{t}$ and
$\beta=I_{0}/I\s{c}$.  These are the incident flux reduced by fluxes $I_t$ or $I_c$ characteristic of \trans or \cis molecules respectively.  We refer to $\alpha$ and $\beta$ subsequently as reduced incident fluxes.
 Ignoring for the moment attenuation, $\alpha$ measures how much a beam intensity $I_0$
leads to \trans conversion, eqn~(\ref{eq:dynamics}), by comparing
$I_0$ to $I\s{t}$, that is the ratio of the forward rate  to the
thermal backward rate, $\alpha = I_0/I\s{t} =
I_0\Gamma\s{t}/(1/\tau)$. Likewise, $\beta$ is the ratio of the
induced to the thermal back rates.

 In terms of the
reduced incident intensities $\alpha$ and $\beta$, the steady state
\textit{trans} and \textit{cis} populations are given by:
\begin{equation}
\begin{array}{ccc}
n\s{t}(x)=\frac{1+\beta
\cI}{1+(\alpha+\beta)\cI}&;&n\s{c}(x)=\frac{\alpha
\cI}{1+(\alpha+\beta)\cI}.
\end{array}\label{eq:ss}
\end{equation}
Here $\cI$ is just $\cI(x)$ since we have the equilibrium case. In
the Eisenbach experiments \cite{Eisenbach:80} the average conversion
was $n\s{c} \sim 0.84$. His measurements of attenuation suggested
$\beta \sim 0$, and thus one can conclude from eqn~(\ref{eq:ss})
that his  $\alpha \sim 5$. Note that $\alpha$ and $\beta$ are
independent of chromophore concentration, but do depend on the
choice of the light polarization \cite{Statman:03}. Experimentally
it is easiest to determine $\alpha$ for a system dilute in
chromophores, where one can ignore the complications arising when
attenuation is significant. These estimates for $\alpha$ are lower
bounds on actual values; including the effects of attenuation
through the cantilever will lead to higher values of $\alpha$. In
later work \cite{Finkphoto} one can deduce that $\alpha \sim 0.8$.

The divergence of the Poynting flux,
 $\frac{\partial I}{\partial x}$, at any
point through the cantilever is equal to the amount of energy taken
out of the beam per unit volume per unit time.  For simplicity, we
ignore curvature leading to obliquity factors for the intensity of
light falling on the surface. That is, we consider small deflections
or  diffuse light.  Energy is taken out of the beam both by the
optically induced \textit{trans}$\rightarrow$\textit{cis} reaction
and by the stimulated \textit{cis}$\rightarrow$\textit{trans}
back-reaction, terms (i) and (ii) above.  The divergence of the
Poynting flux is thus related to the sum of the rates of these two
processes. Thus:
\begin{eqnarray}
\frac{\partial I}{\partial x} =
-\gamma\s{t}\Gamma\s{t}I(x,t)n\s{t}(x,t)-\gamma\s{c}\Gamma\s{c}I(x,t)n\s{c}(x,t)\label{eq:atten}\end{eqnarray}
 where  the constant $\gamma$ in each case subsumes the energy of an incident photon, $\hbar\omega$, the reciprocal of the quantum efficiency $\Phi$ for the relevant
 transition, and the absolute number density of chromophores,
 $\rho_{ph}$ (which could differ between \trans and \cis forms since bent molecules pack less efficiently).  The appearance of $\Phi$ as an inverse is required
 since for each successful transition in the rate $\Gamma\s{i}I
 n\s{i}$ (i = t,c) there will be unsuccessful absorptions that do
 not contribute to $\partial n\s{t}/\partial t$ in
 (\ref{eq:dynamics}), but nevertheless still deplete the optical
 beam and contribute to $\partial I/\partial x$ in (\ref{eq:atten}).

Eqns~(\ref{eq:dynamics}) and (\ref{eq:atten}) are a pair of coupled,
non-linear, first order partial differential equations for $I(x,t)$
and $n\s{t}(x,t)$. Solving these equations subject to the boundary
conditions $I(0,t)=I\s{0}$ and $n\s{t}(x,0)=1$ is in general
complex, though analytically possible in some limits.   We return to
this  problem elsewhere \cite{CorbettDyn:08} and here take the time-independent, equilibrium state. Thus,
using $n\s{c} = 1 - n\s{t}$, dividing through by the incident
intensity $I_0$, and letting $d\s{t} =
 1/(\gamma\s{t}\Gamma\s{t})$ and $d\s{c} =
 1/(\gamma\s{c}\Gamma\s{c})$ denote the characteristic lengths for optical attenuation by \textit{trans}
 and \textit{cis} chromophores respectively, one obtains
 \cite{Statman:03,Corbett:07}
\begin{eqnarray}\frac{d \cI}{d x} = -\left(\left[\frac{1}{d\s{t}}-\frac{1}{d\s{c}}\right]n\s{t}+\frac{1}{d\s{c}}\right)\cI(x).\label{eq:atten_lengths}
\end{eqnarray}
  In terms of the parameters $\alpha$
and $\beta$, the ratio of the \trans and \cis lengths is
$d\s{t}/d\s{c}=(\gamma\s{c}/\gamma\s{t})(\beta/\alpha)$. The ratio
$\eta = \gamma\s{c}/\gamma\s{t}$ is the ratio of the quantum
efficiencies for the $\trans\rightarrow\cis$ and $\cis \rightarrow
\trans$ reactions, that is $\eta=\Phi_{tc}/\Phi_{ct}$. We shall take
$\eta = 1$ in the numerical illustrations in this paper.
 We have ignored any attenuation by the host material; one
could include such effects by adding a simple Lambert-Beer term
$-I(x,t)/d\s{h}$ to the RHS of eqn~(\ref{eq:atten}) or
eqn~(\ref{eq:atten_lengths}).See section \ref{subsect:BeerLambert} for an analysis.

 Inserting the steady-state expression for $n\s{t}$ from
eqn~(\ref{eq:ss}) into eqn~(\ref{eq:atten_lengths}), and then
integrating w.r.t. $x$, subject to $\cI(x=0)=1$, we obtain (see also \cite{Statman:03}):
\begin{equation}
\ln \cI+\left(\frac{\alpha-\eta\beta}{\beta'
}\right)\ln\left(\frac{1+\beta' \cI}{1+\beta'
}\right)=-\frac{x}{d\s{t}},\label{eq:intensity}
\end{equation}
where $\beta'=\beta(1+\eta)$. In general this expression is very
different from Beer's Law, $\cI(x) = \exp(-x/d)$.

Deviations from Beer's Law come about because at high
intensities the \cis population increases (bleaching) and is
generally less absorbing than the \trans species.  Optical
penetration is then more effective and is of great significance for
photo-mechanics.  To most simply see how non-linearities (bleaching) manifest
themselves, consider the limit $\beta\rightarrow 0$
\cite{Corbett:07} that arises when stimulated \cis back-conversion
is weak (for instance in the work of Eisenbach).  Under those
circumstances $\Gamma\s{c} \sim 0$ in eqn~(\ref{eq:dynamics}), and
then  (\ref{eq:atten_lengths}) takes the form:
\begin{eqnarray}\frac{d \cI}{d x} = -\frac{n\s{t}}{d\s{t}}\cI(x).\label{eq:atten_beta0}
\end{eqnarray}
A non-Beer form arises because $n\s{t}= 1/(1+\alpha\cI)$ itself
depends on $\cI$, eqn~(\ref{eq:ss}).  Integration gives
 \begin{eqnarray}\ln \cI +\alpha(\cI-1)=-x/d\s{t},\label{eq:intensityprl}
\end{eqnarray}
also a $\beta \rightarrow 0$ limit of (\ref{eq:intensity}).  A formal solution
of which is $\cI(x)=\frac{1}{\alpha} W\s{L}(\alpha e^{\alpha-\frac{x}{d\s{t}}})$, where $W\s{L}(x)$ is the Lambert-W function~\cite{lambert}. The
non-exponential decay persists until around  $\alpha\cI < 1$,
whereupon $n\s{t}$ becomes independent of $\cI$ and
(\ref{eq:atten_beta0}) reverts to simple exponential form.

The limiting cases of absorption are important.
 \nl (i) $(\alpha + \beta)\cI \ll 1$.  Now $n\s{t} \approx 1$ and
 $n\s{c} \approx 0$ which renders (\ref{eq:atten_lengths}) trivially
 of the Beer form.  The limit obtains when $\alpha = I_0/I\s{t}$ and
 $\beta = I_0/I\s{c}$ are both $\ll 1$ (since $\cI$ is bounded by 1), that is, the incident beam is weak compared with
 both material fluxes $I\s{t}$ and $I\s{c}$.  It also obtains when $\alpha$
 and $\beta$ are not small, but when $\cI \ll 1/(\alpha + \beta)$,
 that is when the beam has diminished (albeit linearly rather than
 exponentially, see the sketch below and also the high intensity traces of Fig.~\ref{fig:intensity0.02}) to the point that
 $n\s{t}= \; {\rm const}\; \approx 1$  and Beer behavior is finally recovered.
 From (\ref{eq:intensityprl}) one sees in fact $\cI(x)
\sim \exp[-(x-d\s{t}\alpha)/d\s{t}]$ for $x > d\s{t} \alpha$, a
shifted Beer form.  Exponential decay is the ultimate fate of all optical beams
provided that cantilevers
 are thick enough to get this diminution of intensity.
\nl (ii)  The high flux limit $\alpha \cI \gg 1$ (with $\beta \sim
0$) is where the forward reaction dominates over thermal back
reaction.  Photo-equilibrium is highly biased away from \trans, that
is in eqn~(\ref{eq:ss}) $n\s{t} \sim 1/(\alpha \cI)$.
Eqn~(\ref{eq:atten_beta0}) reduces to $\cI' \sim - 1/(\alpha
d\s{t})$, whence $\cI \sim 1 - x/(\alpha d\s{t})$.  This linear
penetration for $x \lesssim \alpha d\s{t} $ is evident in
Fig.~\ref{fig:intensity0.02} at the higher $\alpha$ values  and is
at the heart of why even highly absorbing systems can be responsive.
When $x \gtrsim \alpha d\s{t} $, decay is again exponential, see
above.
\nl (iii)  The high flux limit, $\beta \cI \gg 1 $, that is
 $ I \Gamma\s{c} \gg 1 / \tau $, is where the stimulated back-reaction
dominates over the thermal back-reaction.  In that case $n\s{t}
\rightarrow \beta/(\alpha + \beta)$ and $n\s{c} \rightarrow
\alpha/(\alpha + \beta)$ are again constants and again
(\ref{eq:atten}) takes a Beer form in the equilibrium limit:
 \be
d \cI/d x = -\cI \beta(1+\eta)/(d\s{t}(\alpha +
\beta)).\label{eq:Beer-larger} \ee
  The decay is exponential, $\cI =
\exp(-x/d\s{eff})$ with an effective decay length $d\s{eff} =
d\s{t}(\alpha +\beta)/(\beta(1+\eta))$. \nl Thus, except for $\beta
\ll 1$, profiles start in a Beer-manner, have an intermediate
non-exponential behavior if we have $\alpha > 1$ with $\beta \ll
\alpha$, and then conclude with another Beer decay.
 The intermediate regime,  $\beta\cI \sim 1$, is where the
stimulated back reaction rate is comparable to the thermal rate. \nl
(iv) When the decay lengths accidentally coincide, $d\s{t} = d\s{c}=
d$, that is $\gamma\s{t}\Gamma\s{t} = \gamma\s{c}\Gamma\s{c}$, one
can easily see in either (\ref{eq:atten}) or in
(\ref{eq:atten_lengths}) that a Beer form pertains at all
intensities or depths into the photoisomerizing medium: $\cI =
\exp(-x/d)$.

Statman and Janossy~\cite{Statman:03}  investigated
photo-isomerization of solutions of the commercially available
azodye Disperse Orange (DO3).  They obtained a ratio of
$\alpha/\beta \sim 5$  while the accessible range of $\alpha$ is
up-to $\sim 80$ (corresponding to an incident flux of
15mW/mm$^{2}$).
The spectral bands for \trans$\rightarrow$\cis and
\cis$\rightarrow$\trans overlap quite strongly for D03; one might
thus expect much smaller ratios of $\beta/\alpha$ to be accessible
when using dyes with more separated absorption bands.  Indeed,
Eisenbach's attenuation study showed that, for his systems, $d\s{c}
\gg d\s{t}$ and thus that $\alpha \gg \beta$, possibly $\alpha \sim
100 \times \beta$.  We thus show results initially for high
ratios $\alpha/\beta$ for a range of incident intensities $\alpha$
and then look at smaller ratios where the non-linear region is not
so pronounced.

As can be seen from eqn~(\ref{eq:ss}), the \cis population is always
reduced when $\beta$ is finite.  Thus for non-zero $\beta$ we
require a larger value of $\alpha$ to achieve  a particular value of
$n\s{c}$. Similarly attenuation will lead to reduced intensities
lower than unity in eqn~(\ref{eq:ss}), again requiring larger values
of $\alpha$ to achieve the same \cis concentration.  Thus estimates
of $\alpha$ from absorption are lower bounds on true values.

\subsection{CASE 1 - $\alpha/\beta=\infty$}
We recall  the case  in which the illuminating light doesn't
excite the \cis$\rightarrow$\trans back-reaction at all, i.e.
$\beta=0$. Thus the intensity is described by
eqn~(\ref{eq:intensityprl}). Plots are given in \cite{Corbett:07}, but Fig.~\ref{fig:intensity0.02} for the case $\alpha/\beta = 50$ of relatively small $\beta$ also shows how for
low $\alpha$ the decay from the surface intensity is exponential,
but that penetration is much deeper for  higher $\alpha$, being
initially a linear decay until finally decaying as an off-set
exponential  beyond the characteristic depth $d\s{t}$, that is for
$x > d\s{t} \alpha$.

Such non-exponential behavior suggests caution when trying to
establish an extinction length from the attenuation of a light beam
on traversing a cantilever.  For instance \cite{Kondo:06} an
attenuation of 99\% on traversing a cantilever of thickness $1\mu$m,
were Beer's Law being followed, would result from an extinction
length of $d\s{Beer} = 1 \mu$m$ /(2\ln(10))\sim 0.22\mu{\rm m}$. But
if $\alpha$ is not small,  much more light penetrates to $x = w$
(Fig.~\ref{fig:intensity0.02} is a guide).  The $d\s{Beer}$ value derived is a
gross over-estimate of $d\s{t}$.  Solving for $d\s{t}$ from
eqn~(\ref{eq:intensityprl})  for a given attenuation $\cI(w)$ on
reaching the back face at $x=w$ yields for 99\% attenuation: $d \sim
w/(\alpha + 4.6) = 1\mu {\rm m}/(\alpha + 4.6) \rightarrow 0.04
\mu{\rm m}$ for $\alpha = 20$. The true $d\s{t}$ associated with a
possible exponential decay may thus be much lower than the value
$d\s{Beer}$ estimated as above, an indication that light has
penetrated much further into the sample than would be expected for a
simple exponential profile. Quantitative measurements of light
attenuation varying with thickness or varying with incident
intensity would resolve this ambiguity about $d\s{t}$ and also allow
a determination of $\alpha$. Attenuation  at one thickness can only
give an upper bound on $d\s{t}$.

The  reasons for departures from Beer's law for the intensity can be
seen by returning the solution $\cI(x)$ to eqn~(\ref{eq:ss}) to obtain the
spatial variation of the \textit{cis} concentration. Fig.~\ref{fig:cis0.02} for the case $\alpha/\beta = 50$ is a guide to the $\beta = 0$ plots (see \cite{Corbett:07}) and displays exponential
decay in $n\s{c}(x)$ following $\cI(x)$ for low intensity ($\alpha =
0.1,\; 0.5$). High incident intensity not only lifts $n\s{c}(x=0)$
at the surface, but also flattens the decay with depth -- high
$n\s{c}$ means low $n\s{t} = 1 - n\s{c}$ and hence fewer
\textit{trans} dye molecules in a state to deplete the incoming beam
(a photo-bleached state). With photo-bleached surface layers,
radiation penetrates well beyond $x \sim d\s{t}$, and equally,
contraction extends deep into the bulk, certainly beyond the Beer
penetration depth $d\s{t}$.

For $\alpha = 0.5$ a point of inflection first appears at the
surface, and moves inwards with increasing incident intensity
$\alpha$. The \textit{cis} fraction at the inflection is always
$n\s{c} = 1/3$ in this model.  In the general case where $\beta \neq 0$ we find
that the value of $\alpha$ at which the point of inflection first appears at the front surface is
a complicated function of the constant ratio $\beta/\alpha$, and the value of the \cis fraction at the inflection point is no longer $1/3$. From eqn~(\ref{eq:ss}), the surface
\textit{cis} concentration is $n\s{c}(0) = \alpha/(1 + \alpha)$,
since there $\cI = 1$, and rises to saturation, $n\s{c} = 1$, as
intensity increases. The precise form of the cantilever bend depends
critically on the shape of these $n\s{c}(x)$ curves.  In particular
the development of a point of inflection allows three intercepts of
the straight, geometric strain curves, and thus three neutral
surfaces, we will later see.  Since $n\s{c}(x)$ changes shape with
increasing intensity, we will find an elastic response that is
highly non-linear with intensity.

\subsection{CASE 2 - $\alpha/\beta=50$}
\begin{figure}[t]
\centering\resizebox{0.45\textwidth}{!}{\includegraphics{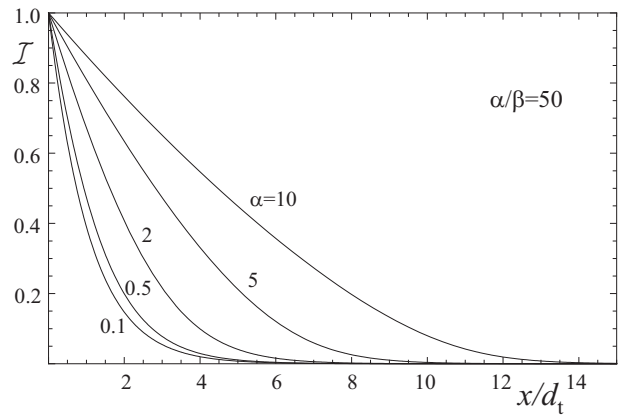}}
\caption[]{The decay in reduced light intensity with reduced depth
for various reduced incident intensities $\alpha$, with a weakly
stimulated back reaction such that $\alpha/\beta=50$. The plots are
similar to those for $\beta = 0$, with small differences we discuss when considering heat generation.}
\label{fig:intensity0.02}
\end{figure}

We now consider the case with a weakly stimulated back-reaction,
$\beta=\alpha/50$, and henceforth take $\eta = 1$.
Fig~\ref{fig:intensity0.02} shows plots of the reduced intensity as
a function of $x/d\s{t}$.  The reduced intensity curves are largely
identical to the $\beta = 0$ case.  Once again,
for small values of $\alpha$ (=0.1, 0.5) the decay is essentially
exponential, with a characteristic length given by $d\s{t}$.
Increasing $\alpha$ leads to deeper penetration, with the initial
decay being essentially linear. However close inspection of the
$\alpha=10$ curve reveals a slight upwards curvature, a result of
the higher order corrections in $\beta$ that take us from
(\ref{eq:intensityprl}) to (\ref{eq:intensity}).  Eventually
$\cI(x)$  becomes exponential at penetration depths $x \sim d\s{t}
\alpha$ significantly greater than $d\s{t}$.

As we show above, for $\beta \gtrsim 1$, that is here
$\alpha \gtrsim 50$, the initial behavior should (briefly) revert to being
exponential before attaining the linear decay associated with
non-Beer. This point is easier to display below when we consider
smaller $\alpha/\beta$ ratios.

\begin{figure}[t]
\centering\resizebox{0.45\textwidth}{!}{\includegraphics{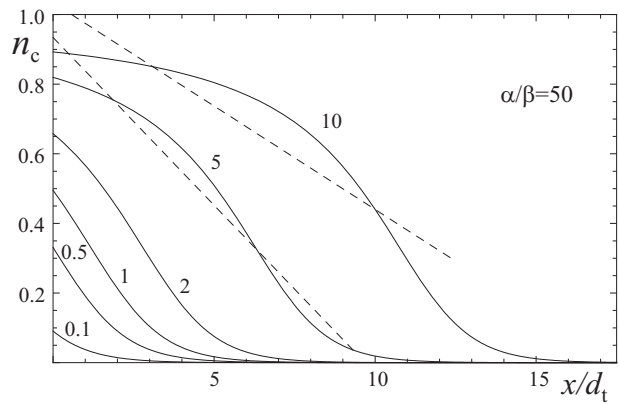}}
\caption[]{The decay in \cis number fraction $n\s{c}(x)$ with
reduced depth for various reduced incident intensities $\alpha$,
with $\alpha/\beta=50$. Increasing
 $\alpha$ extends the conversion to \cis
 to greater depths because of photobleaching of the surface layers. Reduced geometric bend strain for
$\alpha = 5,\, 10$ are shown as straight dotted lines. For
cantilevers of thickness $w =w* = 9.313d$ and $w=12.5d$
respectively, there are three and two intersections (neutral
surfaces) with the photo-strain curves.} \label{fig:cis0.02}
\end{figure}

The \cis profiles as a function of depth are shown in
Fig~\ref{fig:cis0.02}.  For the values of $\alpha$ plotted, the curves
are essentially identical to those in the $\beta = 0$ case.  High incident intensities result in
larger \cis fractions near the surface, and a flatter decay as
before. The point of inflection occurs for a \cis fraction less than the 1/3 found in the $\beta = 0$ case, but it is still of importance that inflections exist for the character and number of neutral surfaces we explore later.

\subsection{CASE 3 - $\alpha/\beta=5$}
\begin{figure}[t]
\centering\resizebox{0.45\textwidth}{!}{\includegraphics{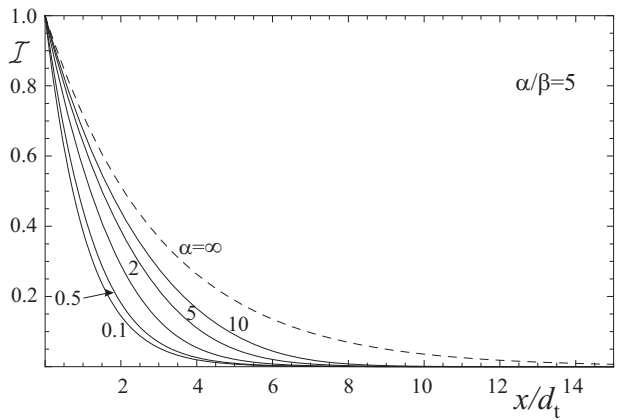}}
\caption[]{The decay in reduced light intensity with reduced depth
for various reduced incident intensities $\alpha$ with
$\alpha/\beta=5$. } \label{fig:intensity0.2}
\end{figure}

Increasing the stimulated back reaction further we take
$\alpha/\beta=5$. The curves for the reduced intensity as a function
of depth, see Fig~\ref{fig:intensity0.2}, now differ somewhat from
those in the small $\beta$ limit, e.g. as in Fig~\ref{fig:intensity0.02}.  For the smaller values of
$\alpha$ the curves remain exponential with a characteristic length
$d\s{t}$. Increasing $\alpha$ leads to some increased penetration,
without showing the long linear decay in reduced intensity seen in
Fig~\ref{fig:intensity0.02}. For $\alpha =
10$ we have $\beta\cI = 2$ at most, a value evidently insufficient
to satisfy limit (iii).  We do not have a finite initial region for
small $x$ where the decay is exponential with $d\s{eff}$ given in
and below eqn~(\ref{eq:Beer-larger}).  For these values of $\alpha$
and $\beta$ one would have   $d\s{eff}=3d\s{t}$. The dashed line
shows the infinite $\alpha$ limit, and corresponds throughout its
range to eqn~(\ref{eq:Beer-larger}), i.e. an exponential with
characteristic length $d\s{eff} = d\s{t} (\alpha
+\beta)/(\beta(1+\eta)) = 3d\s{t}$.  Note that the initial of the
$\alpha = 10$ curve is close to that of the $\alpha = \infty$ curve.

\begin{figure}[t]
\centering\resizebox{0.45\textwidth}{!}{\includegraphics{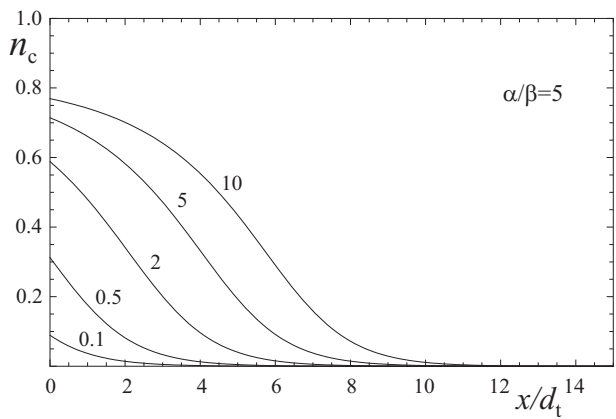}}
\caption[]{The decay in \cis number fraction $n\s{c}(x)$ with
reduced depth for various reduced incident intensities $\alpha$ with
$\alpha/\beta=5$. } \label{fig:intensity_cis3}
\end{figure}

\section{Other absorption processes}\label{sect:other}
At least two other influences are important in non-linear
absorption.  How does one deal with the absorption of
the host material that is not a chromophore.  In principle this is
the simplest form of absorption and for weak beams where both processes are Beer-like the effect can simply be
divided out. We examine how this procedure translates to the non-linear case. Another, more complex effect is that of the host
medium when it is a solid.  The isomerisation processes can be
biased by mechanical effects, as is well-known from linear
experiments.  These influence the non-linear absorption for intense
beams.

\subsection{Beer-Lambert host absorption}\label{subsect:BeerLambert}
 It will always be the case
that the  host for the chromophores will also absorb
light. For simplicity ignoring the backreaction, $\beta = 0$, the
absorption equation (\ref{eq:atten_lengths}) becomes
\begin{eqnarray}\frac{d \cI}{d x} = -\left(\frac{n\s{t}}{d\s{t}}+\frac{1}{d\s{h}}\right)\cI(x).\label{eq:atten-bg}
\end{eqnarray}
The absorption length $d\s{t}$ of course emerges naturally by
measuring the absorption of a sample without dye, $-\ln(\cI_w) =
w/d\s{h}$. Using $n\s{t} =
1/(1+\alpha \cI)$, and denoting the ratio of absorption lengths
$\psi = d\s{t}/d\s{h}$, integration yields: \begin{eqnarray} \ln(\cI_w) + \frac{1}{\psi} \ln
\left(\frac{1 + \psi(1 + \alpha \cI_w)}{1 + \psi(1 + \alpha)}
\right) = - w/d\s{eff} \label{eq:profile-bg}
\end{eqnarray}
where the effective absorption length is $1/d\s{eff} = 1/d\s{t} +
1/d\s{h}$.  The relation (\ref{eq:profile-bg}) for $\cI_w$ has
close similarities to that for the profile with a back reaction,
eqn~(\ref{eq:intensity}).   The limit $\psi \rightarrow 0$ of eqn~(\ref{eq:profile-bg}) of no
host absorption leads to the previous Lambert-W expression
given below eqn~(\ref{eq:intensityprl}).  In the conclusions we discuss how host absorption can be allowed for in measurements.

\subsection{Absorption of intense beams in solid hosts}\label{subsect:glasses}
The role of the host has been much studied in dynamical studies of
absorption.  When the host, typically a polymer or a network, is
below the glass temperature then it is observed that there is more
than one characteristic decay time for the \cis population of guest
chromophores.  Eisenbach \cite{Eisenbach:78} clearly observed two
times, the relative weight of the processes being dependent upon
temperature. The slower process was analogous to that observed in a
liquid host and becomes the only process present at elevated
temperatures.  The faster process was speculated to arise from the
strained state of a \cis isomer; a fraction of the molecules when in
a bent state are in conflict with the solid matrix around them and
thereby decay more quickly to the \trans form.  Other authors
\cite{Kryszewski:80} suggest that there is a spectrum of relaxation
times corresponding to a range of environments that \cis
chromophores find themselves in.   Eisenbach also found that, when
above its glass temperature, rubber provides a liquid-like
environment for its chromophore guests.  There was only one
relaxation time and this was comparable to that found when
chromophores were dissolved in liquids, a fact of considerable
importance since elastomers are used as photo-mechanical media.

Ignoring for simplicity photo-induced back reactions (as shown to be
the case for instance in the Eisenbach study), the dynamical
equation (\ref{eq:dynamics}) is modified in that in principle the
forward rate $\Gamma\s{t}$ could become a function its environment
(denoted by $p$) and the thermal back reaction rate $1/\tau$ is
certainly a function of $p$.  The result is that the characteristic
reduced intensity is now a function of environment and we denote it
by $\alpha_p = \Gamma\s{t}^p\tau^pI_0$.  The equilibrium \trans
number fraction for chromophores in the environment $p$ is thus
$n\s{t}^p = 1/(1 + \alpha_p \cI)$, see the original equation
(\ref{eq:ss}).  Let the probabilities of finding a chromophore in
the $p$ environment be $a_p$.  Then the total \trans volume fraction
of chromophores is $n\s{t} = \sum_p a_p/(1 + \alpha_p \cI)$, and it
is this $n\s{t}$ that must appear in eqn~(\ref{eq:atten}):
\begin{eqnarray}
\frac{1}{\cI} \frac{d\cI}{dx} &=& - \frac{1}{d\s{t}}\sum_p
\frac{a_p}{1 + \alpha_p \cI} \\
-\frac{w}{d\s{t}} &=& \int_1^{\cI} \frac{d\cI}{\cI \left(\frac{a}{1+
\alpha_1\cI} + \frac{1- a}{1+ \alpha_2\cI}\right)}\\
-\frac{w}{d\s{t}} &=& \frac{1}{\alpha}\left(\alpha_1 + \alpha_2 -
\alpha - \frac{\alpha_1\alpha_2}{\alpha}\right)\ln\left(\frac{1 +
\alpha \cI}{1+ \alpha}\right) + \nonumber \\ && + \ln(\cI) - (1-\cI)
\frac{\alpha_1\alpha_2}{\alpha}
\end{eqnarray}
The first equation is general, the second two have been specialised
to the Eisenbach case of two decay rates and thus $\alpha_1$ and
$\alpha_2$ with weights $a$ and $1-a$ respectively.  Here $\alpha =
a \alpha_1 + (1-a)\alpha_2$; note that this is not an averaged
$\alpha_p$.  Again, this result has similarities of form with the
$\beta \neq 0$ form (\ref{eq:intensity}) and the host absorption result
(\ref{eq:profile-bg}), and is derived by the same kinds of
integrations.

\section{Optically-induced curvature}\label{sect:curvature}
Fig.~\ref{fig:slender} shows a cantilever with radius of curvature
$R$.  The geometric strain from bending is $x/R + K$, where $R$ is
the radius of curvature adopted by the cantilever and $K$ is a mean
strain, both to be determined for a given thickness $w$ and
illumination. Illumination changes the natural length of the sample,
the actual strain with respect to this new natural length is
$x/R~+~K~-~\epsilon\s{p}$ which, if we further reduce $x/R$ and $K$
by the dimensionless constant $-A$ connecting photo-strain
$\epsilon\s{p}$ and the \textit{cis} concentration, we obtain $x/R +
K + n\s{c}(x)$ for the effective reduced strain.  A linear relation
between \cis concentration and strain is probably valid for nematic
glasses, but for elastomers it is possible to reach the isotropic
state by illumination at temperatures close enough to the
nematic-isotropic transition and the relation is no longer linear,
but can be mapped on to the observed variation of strain with
temperature\cite{Finkphoto}. The mechanical stress $\sigma$ is
related linearly to the strain via the Young's modulus $E$. Since
there are no external forces nor external torques, mechanical
equilibrium requires vanishing total force and moment across a
section, thus:
 \begin{eqnarray}
 \!\int_{0}^{w}\sigma(x)dx&=& E\int_0^w \!\!\left(\frac{x}{R} + K + n\s{c}(x)\right)\!dx=0,\nonumber\\
\int_{0}^{w}x\sigma(x)dx&=&\!\!E\int_0^w \!\!\! x\left(  \frac{x}{R} + K + n\s{c}(x)\right)\!dx=0.  \label{eq:equilib}
\end{eqnarray}
 When the modulus is constant it
 drops out, but must generally be retained  (for some photo-glasses $E$ is known to vary with
illumination \cite{vOosten:07}).  Performing these integrations we
have:
\begin{eqnarray}
\frac{w^2}{2R}+Kw&=&-\int_{0}^{w}n\s{c}(x)dx\label{eq:intone}\\
\frac{w^3}{3R}+\frac{Kw^{2}}{2}&=&-\int_{0}^{w}xn\s{c}(x)dx\label{eq:inttwo}.
\end{eqnarray}
Simplifying between these two expression we obtain for the radius of curvature:
\begin{equation}
\frac{1}{R}=\frac{12}{w^{3}}\int_{0}^{w}\left(\frac{w}{2}-x\right)n\s{c}(x)dx\label{eq:intthree}
\end{equation}
Eqn~(\ref{eq:atten_lengths}) can be rearranged to give an expression
for $n\s{c}(x)$, recalling $d\s{t}/d\s{c} = \eta \beta/\alpha$,
$\eta=\gamma\s{c}/\gamma\s{t}$, and $n\s{t} = 1 - n\s{c}$:
\begin{equation}
n\s{c}(x)=\frac{1}{1-\eta\frac{\beta}{\alpha}}+\frac{d\s{t}}{1-\eta\frac{\beta}{\alpha}}\frac{1}{\cI}\frac{d
\cI}{d x},
\end{equation}
Inserting this expression into eqn~(\ref{eq:intthree}) and changing
integration variables $\int_0^wdx \rightarrow \int_{\cI_0 =
1}^{\cI_w}d\cI$ we have:
\begin{equation}
\frac{d\s{t}}{R}=12\left(\frac{d\s{t}}{w}\right)^{3}\frac{1}{1-\eta\frac{\beta}{\alpha}}
\int_{1}^{\cI_{w}}\left(\frac{w}{2d\s{t}}-\frac{x}{d\s{t}}\right)\frac{d\cI}{\cI}.
\end{equation}
Substituting for $x/d\s{t}$ and $w/d\s{t}$ from
eqn~(\ref{eq:intensity}) and integrating, we
 obtain (with $\beta' = \beta(1+\eta))$:
\begin{eqnarray}
\frac{d\s{t}}{R}&=&\frac{12\alpha}{\beta'\left(w/d\s{t}\right)^{3}}\Big{[}Li_{2}(-\beta')-Li_{2}(-\beta'\cI_{w}) - \nonumber\\
&&-\half\ln(\cI_{w})\ln\left[(1+\beta'\cI_{w})(1+\beta')\right]\Big{]}
\end{eqnarray}
where $Li_{2}(x)=\int^0_x dt\frac{\ln(1-t)}{t} =
\sum_{k=1}^{\infty}\frac{x^{k}}{k^{2}}$ is the dilogarithm
\cite{Abramowitz:72}. The limit $\beta\rightarrow 0$ within this
expression  recovers our earlier expression for the curvature
\cite{Corbett:07}:
\begin{eqnarray}
\frac{d\s{t}}{R}&=& \frac{12\alpha d\s{t}^3}{ w^3} \times\label{eq:prl} \\
&&\times \left[\frac{w}{d\s{t}}\cI_w - (1 -
\cI_w)\left(1-\frac{w}{2d\s{t}}\right) - \frac{\alpha}{2}(1 -
\cI^2_w) \right] \nonumber
\end{eqnarray}


At low incident light intensity, $\alpha \rightarrow 0$,  analysis
\cite{Mahadevan04} for exponential decay gave maximal reduced
curvature $w/\alpha R$ for $w/d \sim 2.63$. In this limit $1/R \sim
\alpha \sim I\s{o}$, hence the division by $\alpha$ to obtain
results universal for all (low) intensities of incident light.  As
intensity increases, the maximum in $w/\alpha R$ moves to larger
$w/d$ because the radiation penetrates more deeply.

\begin{figure}[t]
\centering\resizebox{0.45\textwidth}{!}{\includegraphics{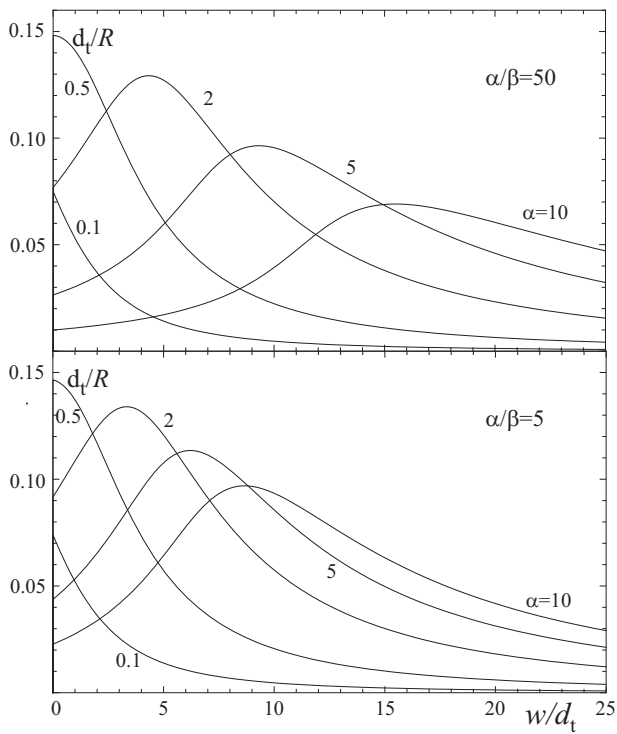}}
\caption[]{Curvature reduced by $1/d\s{t}$ against reduced
cantilever thickness $w/d\s{t}$ for various incident reduced light
intensities $\alpha$, with   $\alpha/\beta = 50$ (upper) and $=5$ (lower).  For high intensities,
bleached surface layers let light penetrate more deeply and hence a
significant fraction of the cantilever has its natural length
contracted.  Bend occurs even for cantilevers much thicker than the
linear penetration depth $d\s{t}$. Maxima in curvature
occur at smaller thicknesses than in the $\beta = 0$ case where back
reaction is purely thermal, and are seen at ever smaller thickness as $\beta$ increases.} \label{fig:nonBeer}
\end{figure}

The non-linear regime, at fixed Beer's Law penetration $d\s{t}$, is
best revealed by  reducing $R$ by $d\s{t}$ instead of by $w$, and by
not reducing $1/R$ by $\alpha$.  Fig.~\ref{fig:nonBeer} plots
$d\s{t}/R$ against $w/d\s{t}$ to reveal maxima in $d/R$ at greater
$w$ as intensity $\alpha$ and thus penetration increases.
  At a given $w/d\s{t}$, one sees
curvature increase initially with intensity, the Beer limit, and
then reduce as penetration increases and gradients of strain are
reduced. Thus appreciable curvature  arises experimentally even in
cantilevers thick in the sense $w \gg d\s{t}$:  in \cite{Ikeda:03}
it appears that curvature is induced even though the cantilevers
involved (with $w = 10 \mu$m) are apparently at least 100 times
thicker than their extinction length! (See also \cite{Kondo:06}.)
The curvatures against thickness for various intensities $\alpha$ in
the case of $\alpha/\beta = 50$ studied above are practically
identical to those in the $\beta = 0$ limit of
\cite{Corbett:07}.  As the back-reaction rate increases
relative to the forward rate, $\alpha/\beta = 5$, the penetration is
less and the curvature maxima move to noticeably smaller reduced
thicknesses $w/d\s{t}$, see lower pane.

In the limit of intense illumination, that is $\alpha \gg 1$, the
intensity profile simplified to $\cI \simeq 1 - x/(\alpha d\s{t})$
for penetration $x < \alpha d\s{t}$ and hence intensity such that
$\alpha \cI > 1$. Under these circumstances $n\s{c} \simeq 1 -
1/(\alpha \cI)^2$ and, importantly for the curvature, $n\s{c}'
\simeq - 1/[d\s{t}(\alpha \cI)^2]$.  If the thickness $w \ll \alpha
d\s{t}$, then the largely linear profile of $\cI(x)$ obtains through
most of the thickness of the cantilever and also $\cI \sim 1$.  One
can then approximate $n\s{c}' \simeq - 1/(d\s{t} \alpha^2)$.  While
the strain follows $n\s{c}(x)$ linearly, then there will be a linear
change of photo-strain, that is new natural state of the cantilever,
through the thickness.  If the geometric bend follows this exactly,
then there is no effective internal stresses in the material.  This
condition fixes the curvature as $d\s{t}/R = A/\alpha^2$. We thus
see that this definition of very strong irradiation involves the
thickness $w$ in relation to $\alpha d\s{t}$. It is a condition that
penetration is so deep that light emerges from the other side of a
heavily dye-loaded sample since there is no region of exponential
decay of $\cI(x)$.  Moreover, because of the deep penetration, the
curvature now  diminishes with increasing intensity (like
$1/\alpha^2$), rather than increasing. The time taken for the
profile to reach the highly bleached form above can be long, a
problem we return to in modeling the non-linear dynamics of intense
absorption \cite{CorbettDyn:08}.

Quantitative measurements of reduced curvature $w/R$ with $I\s{o}$
are required to probe this complex dependence of curvature on
thickness and incident intensity.  Particular care must be taken to
irradiate long enough for equilibrium to be reached, and possibly
curvature to reduce from high intermediate values.

\section{Strain Distributions}\label{sec:neutrals}


In the linear case \cite{Mahadevan04} of bend induced by exponentially decaying optical intensity, two neutral surfaces arise,
that is surfaces of zero stress where the geometric strains arising
from curvature happen to match the local photo-strain: $x\s{n}/R + K
+ n\s{c}(x\s{n}) = 0$.  A classical cantilever bent by imposed
terminal torques has a single neutral surface  at its mid-point
$x_{n}=w/2$, so that stresses that are equal and opposite about the
$x=w/2$ sum to zero to give no net force.  In the current case of
cantilevers bending because of strains generated internally by
light, rather than by external imposition of torques, the additional
constraint of no net torque leads to a more complex distribution of
stresses which gives rise to more than one neutral surface.

The maximum values for both the curvature $1/R$ and contraction $K$
are intimately related to the positions of the neutral surfaces.
Differentiating  eqns~(\ref{eq:intone}) and (\ref{eq:inttwo}) with
respect to the cantilever thickness $w$ and solving between the
resulting equations, we obtain the relationship:
\begin{equation}
\frac{d R}{d w}=\frac{3R^{2}}{w}\frac{d K}{d w},
\end{equation}
thus both $1/R$ and $K$ are maximized for the same cantilever
thickness. Furthermore, returning $d R/d w=
d K/d w = 0 $ to either of the differentiated
equations, one finds that  the thickness at which this happens,
$w^*$, satisfies $w^{*}/R+K+n\s{c}(w^*)=0$, i.e. the straight line
of geometrically induced strain intersects the photo-strain ($\sim
n\s{c}(x)$) curve at the back surface of the cantilever.  The back
surface  is a neutral surface when the curvature is maximized. The
curve $n\s{c}(x)$ is always a strictly decreasing function of $x$;
however there is a fundamental difference between the curves in the $\beta = 0$ case for $\alpha<0.5$ and those for $\alpha>0.5$ --
the latter $n\s{c}(x)$ curves have a point of inflection
($n''\s{c}(x)=0$).  The straight line of geometrically generated
strain imposed from curvature, $-(x/R+K)$, can intersect the
$n\s{c}(x)$ curve at most twice if there is no point of inflection,
and at most three times if there is a single point of inflection. It
is not possible for the strain to satisfy the constraints of
vanishing force and torque, that is satisfy eqns~(\ref{eq:equilib}),
and have a neutral surface at the back surface of the beam with only
two neutral surfaces; thus there is no maximum for the curvature
unless the underlying \cis curve has an inflection point.  For $\beta \ne 0$ the $\alpha$ value that divides the curves into those with and those without an inflection point is no longer exactly $\alpha = 0.5$ but, as can be seen in  Fig~\ref{fig:cis0.02}, for small $\beta$ the division is still at an $\alpha$ close to 0.5.

\begin{figure}[!t]
\centering\resizebox{0.43\textwidth}{!}{\includegraphics{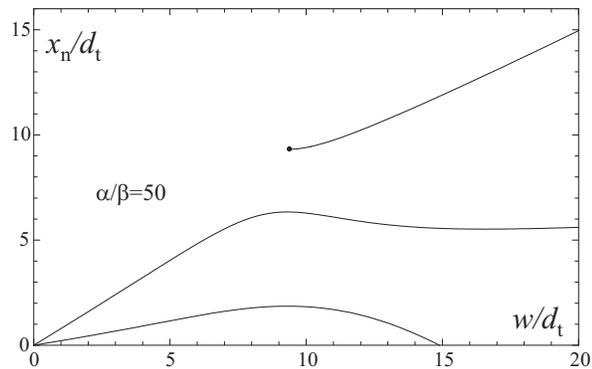}}
\caption[]{Neutral surfaces for fixed incident light intensity
$\alpha =5$, $\alpha/\beta = 50$ as cantilever thickness $w$
changes.  At the $w$ maximizing curvature, a third neutral surface
appears from the rear face; at greater $w$ a neutral surface is lost
at the front face.} \label{fig:neutrals}
\end{figure}

To illustrate the significance of inflections, two sample curvature
strains of purely geometric origin are superimposed in Fig~\ref{fig:cis0.02}.
On the $\alpha=10$, $\beta=0.2$
curve is also plotted the straight line $-(x/R+K)$ for a cantilever
of thickness $w=12.5d\s{t}$, a thickness that is before the
inflection point in the $n\s{c}(x)$ curve.  As can be seen, the
straight line intersects the underlying $n\s{c}(x)$ curve only twice
in satisfying eqns~(\ref{eq:equilib}) and locates only two neutral
surfaces. On the $\alpha=5$, $\beta=0.1$ curve is also plotted
$-(x/R+K)$ for the critical thickness $w=w^{*}=9.313 d\s{t}$ where the third neutral surface first appears. One sees that this
line intersects the underlying $n\s{c}(x)$ curve three times,
internally twice with the third intersection (neutral surface)
coinciding with the back surface. This line is that of maximal
possible slope, $d R/d w = 0$. With further increasing
thickness, $d/R$ decreases.  Eventually a neutral surface migrates
to the front face and is lost.  The cantilever continues to have
only 2 neutral surfaces thereafter.
Fig.~\ref{fig:neutrals} shows how  the neutral surfaces change with
increasing thickness at fixed illumination $\alpha = 5$, $\alpha/\beta = 50$.
 Features are slightly shifted to smaller $w$ in comparison with the $\beta=0$ case of \cite{Corbett:07}.
 With two
regions of compression and two of elongation, one expects subtle
behavior when  considering compression-induced director
rotation \cite{Verduzco:06} in such cantilevers, to which we return
elsewhere.

\section{Temperature effects}\label{sec:temperature}
We have neglected the effect of heat generated by absorption of
light.  Gradients of optical intensity in the cantilever might be
expected to generate gradients of temperature and thus of thermal
contraction, leading to thermal bend. Experiments where polydomain
elastomers bend in the direction of the polarization of light
\cite{Ikeda:03} show directly that optical effects dominate (see the
analysis of polydomains in \cite{Corbett:06})over thermal component
of bend. We here quantify the relative size of optical and thermal
effects.  Let the temperature distribution in the beam be
$\theta(x,t)$, and take the origin of the temperature scale be such
that the ambient temperature outside the cantilever is zero, $\theta
= 0$. The temperature distribution satisfies a continuity equation
in which the heat flux contains the usual thermal gradient term
$-\kappa \frac{\partial \theta}{\partial x}\mathbf{\hat{x}}$ and the
divergence of the Poynting flux, $I(x)\mathbf{\hat{x}}$, is a source
term:
\begin{equation}
C\frac{\partial \theta}{\partial t}-\kappa\frac{\partial^{2}\theta}{\partial x^{2}}=-I_{0}\frac{\partial \cI}{\partial x} \label{eq:heatdiffuse}
\end{equation}
where $C$ is the specific heat of the the cantilever per unit volume
and $\kappa$ is the thermal conductivity perpendicular to the
director, that is along the normal to the flat surface of the
cantilever. The diffusion co-efficient $D$ is given by the ratio
$\kappa/C$. Typical values for elastomers are around $\sim
10^{-7}$m$^{2}$s$^{-1}$ ~\cite{broerman}, with some anisotropy in
$D$ arising in nematic elastomers from anisotropy in the
conductivity that we ignore here since thermal effects will in any
case turn out to be small. The time taken for heat to diffuse across
the thickness of the cantilever is $\approx w^{2}/D \sim 0.001s$ for
a $10\mu$m sample~\cite{Ikeda:03,vanoosten:06}. For times
significantly longer than this, one obtains the steady state
solution $\theta(x)$.  At the front and back surfaces there are
convective losses which are described by Newton's law of cooling,
that is the heat flux carried away from a surface is $\delta
\theta(0)$ (the temperature outside is $\theta = 0$), where $\delta$
is the convective heat transfer co-efficient.  For free convection
in air, $\delta \approx 5$ W m$^{-2}$ K$^{-1}$. These convection
losses are equal to the thermal flux of heat at the respective
surfaces:
\begin{eqnarray}
\theta(0)-\left.\frac{\kappa}{\delta} \frac{d \theta}{d x}\right|_{x=0}&=&0\nonumber\\
\theta(w)+\left.\frac{\kappa}{\delta} \frac{d \theta}{d x}\right|_{x=w}&=&0,
\end{eqnarray}
where the signs reflect the direction of the outward surface normal.
The thermal conductivity is $\kappa \approx 0.2$ W m$^{-1}$ K$^{-1}$
\cite{niu:06}. The solution to eqn~(\ref{eq:heatdiffuse}) satisfying
boundary conditions is:
\begin{eqnarray}
\frac{\theta(x)}{\bar{\theta}}&=&1-\frac{\left(1+\frac{\delta
x}{\kappa}\right)}{\left(2+\frac{\delta
w}{\kappa}\right)}\left[1+\cI_{w}+\frac{\delta
w}{\kappa}\int_{0}^{1}\cI(wy)dy\right]\nonumber\\&&+\frac{\delta
w}{\kappa}\int_{0}^{\frac{x}{w}} \cI(wy) dy \label{eq:temp}
\end{eqnarray}
The characteristic temperature $\bar{\theta}=I_{0}/\delta$ is the
temperature that the sample surface would need to attain in order to
lose by Newton cooling all the heat equivalent to the incident
Poynting energy. Both of the integrals appearing in this expression
have been scaled such that their value is bounded by unity.  The
scale of their contributions is thus set by their pre-factor $\delta
w/\kappa$. This dimensionless quantity compares the thickness of the
sample $w$ with the thermal penetration length set by the boundary
conditions $\kappa/\delta$.

There are several interesting limits to this equation:

(i) Using the values for $\kappa$ and $\delta$ given above and assuming a cantilever thickness $w\sim 10\mu$m we obtain $\frac{\delta w}{\kappa}\approx2.5\times10^{-4}<<1$, and thus the temperature distribution is essentially constant throughout the sample, that is:
\begin{equation}
\frac{\theta(x)}{\bar{\theta}}=\left(\frac{1-\cI_{w}}{2}\right)+O(\delta
w/\kappa)
\end{equation}

A constant temperature through the sample leads to contraction along
the long axis of the cantilever, but it will not induce bending,
since this requires differential contractions.  Therefore in this
limit, which is close to experimental reality, bending effects in
the steady state are due to the optical effects discussed
previously, rather than heating.

(ii) Heat is produced proportionately to the rate optical intensity
diminishes. There are  regimes in which intensity $\cI(x)$ decays linearly
with depth into the cantilever, such as the $\alpha=10$, $\beta = 0$ case of \cite{Corbett:07} where there is linearity throughout if  $w\leq 10d\s{t}$, and here in
Fig~\ref{fig:intensity0.02} for $\alpha=10$, $\beta = 0.2$ where the range of thicknesses where the entire profile is linear is slightly reduced by the influence of stimulated back reactions ($\beta \ne 0$), say $w\leq 8d\s{t}$.  For these conditions,  that is where $d{\cal I}/dx$ = const., then heat is generated at the same rate through the
cantilever. In the steady state it diffuses to the surfaces,
symmetrically if the surfaces are at the same temperature, and hence
heat does not contribute at all to the bending.  One sees this since
now eqn~(\ref{eq:temp}) becomes:
\begin{equation}
\frac{\theta(x)}{\bar{\theta}}=\left(\frac{1-\cI_{w}}{2}\right)\left[1+\frac{\delta
w}{4\kappa}-\frac{\delta}{\kappa
w}\left(x-\frac{w}{2}\right)^{2}\right],
\end{equation}
a temperature distribution which is indeed symmetric about the
mid-point of the cantilever.  Since the thermally-induced strains
are also symmetric, they will not produce bending, although once
again they will produce an overall contraction.  If the cantilever
thickness $w$ is increased beyond the linear profile interval in e.g.
Fig~\ref{fig:intensity0.02}, then asymmetry in the heat production
starts to occur, with less heat generated towards the back face.
However, the magnitude of the asymmetry is reduced by diffusion and
the thermal contribution to bend sets in only slowly with increasing
$w$, see (i) above.

The neglect of heat is thus justified in two limits, firstly the
convective heat losses from the boundary are likely to result in a
uniform temperature distribution through the sample for
experimentally realistic values of the thermal conductivity $\kappa$
and the convective heat transfer co-efficient $\delta$.  Further, in
the regime of linear (i.e. non-exponential) decay of intense beams
there is no thermal component of bend.

Thermal effects become more extreme, especially in elastomers, if an
interior region of the cantilever's temperature exceeds the
nematic-isotropic transition temperature.  At this temperature an
extremely large strain can develop in elastomers, and possibly in
glasses.  If it occurs in a region symmetrically disposed about the
cantilever mid plane, it could lead to pronounced contractions, but still
not to bend.  If it occurs in an asymmetrically disposed region, it
could lead to large bends -- an extreme case we return to in
considering specifically elastomers and thus at the same time
considering director rotation.  Additionally one would there
consider the displacement of neutral planes due to volumes of the
cantilever suffering large contractions of their natural lengths.
\begin{figure}[!b]
\centering\resizebox{0.45\textwidth}{!}{\includegraphics{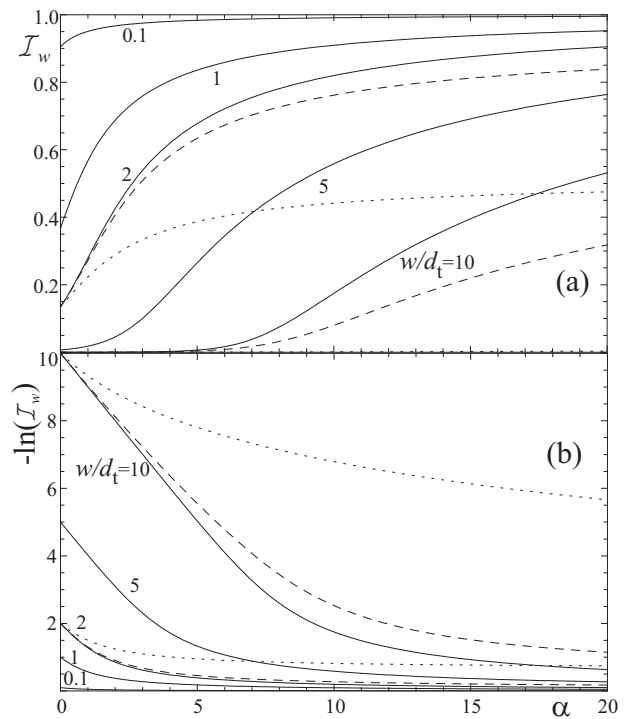}}
\caption[]{(a) Scaled intensity at the back surface $\cI_{w}$ as a
function of the incident intensity $\alpha$ for
various cantilever thicknesses $w/d\s{t}$ with $\beta=0$ (solid lines),
 and with $\alpha/\beta = 50$ (dashed lines) and $\alpha/\beta =5$ (dotted lines) for two marked thicknesses. (b) Intensity plotted in the form of the absorbance
   $-\ln(\cI_{w})$ as a function of $\alpha$ with cantilever
thicknesses and $\beta$ values as in the upper figure.} \label{fig:predict}
\end{figure}

\section{Conclusions}\label{sec:conclusions}
We have shown that non-linear absorption (that is, non-exponential
profiles) can be invoked to explain how bending can arise in
cantilevers where, within Beer's law, one would otherwise expect no
response. At high enough incident light intensities, there can be
photo-bleaching and thereby penetration of radiation and thus
elastic response even in cantilevers so heavily loaded with dye that
the Beer penetration depth of the linear regime is
insignificant compared with the cantilever's thickness. The
non-linear response in the high dye-loaded limit is possibly of the
greatest experimental relevance.  We have explored the roles of optically-stimulated de-excitation of dyes and of optically-generated heat in mechanical processes. Stimulated decay can be very important, but the role of heat seems minor.
\\
\\
An  experiment \cite{Statman:03} to explore the non-linear absorption processes we
have described  measures the intensity emergent at the back surface
for a sample of fixed width $w$ illuminated at the front surface.
Altering the incident intensity is equivalent to varying the
parameter $\alpha$. Taking a fixed depth (the thickness itself) $w/d\s{t}$ in Fig~\ref{fig:intensity0.02} and \ref{fig:intensity0.2} and increasing reduced incident power $\alpha = I_0/I\s{t}$ is to take a slice through these figures to reveal increasing relative penetration, that is an increased  reduced exit intensity $\cI_w(\alpha) = I_w/I_0$, see Fig~\ref{fig:predict}(a).  Scaling the output intensity by its incident
value would produce a horizontal  line $\cI_w = \exp(-w/d\s{t})$ as a function of $\alpha$ for simple
Beer law attenuation; deviations from this line as power increases away from $\alpha = 0$ are thus  signs of
non-Beer attenuation processes.  In thin cantilevers
$w/d\s{t}=0.1$, $\cI_{w}$ is close to unity, and only increases
slowly as $\alpha$ is increased -- for thin cantilevers nearly all of
the incident flux is transmitted in any case.
Conversely for thick cantilevers $w/d\s{t}=10$ we see that there is
very little transmittance for small $\alpha$. For larger incident
intensities, such that $\alpha \gtrsim 5$, the transmittance begins
to increase rapidly with $\alpha$ - a consequence of the increased
penetration due to non-linear absorption processes.  For two particular sample thicknesses, $\beta \ne 0$ cases are shown too.  The curves show that a finite induced back reaction rate reduces enhanced penetration, as one would expect. For the thickest sample, the higher $\beta$ value is sufficient to effectively eliminate penetration of the beam. Deviation from Beer absorbance is perhaps better seen from plots of non-linear absorbance itself, see Fig.~\ref{fig:predict}(b).  An experimental example, and its relation to their equivalent of our eqn~(\ref{eq:intensity}), is discussed at length by Statman and Janossy \cite{Statman:03}.

\begin{figure}[!t]
\centering\resizebox{0.45\textwidth}{!}{\includegraphics{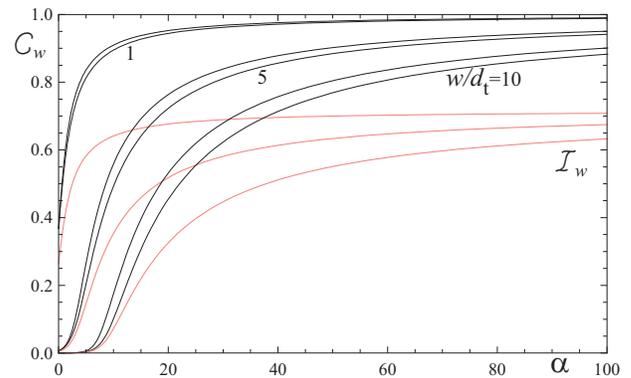}}
\caption[]{Corrected intensity at the back surface $\cC_{w}$ as a
function of the incident intensity $\alpha$ for
various cantilever thicknesses $w/d\s{t}$.  In each case the upper curve is without host absorption, the lower with rather heavy host absorption such that $w/d\s{h} = 0.3$.  For comparison, lower lighter lines, the corresponding uncorrected $\cI_w$ curves are shown.} \label{fig:Casper}
\end{figure}

Despite host and dye absorptive processes being mixed in a non-linear differential equation, an approximate allowance by division by the host transmittance seems to work well:  $\cC_w = \cI_w/{\rm e}^{-w/d\s{h}}$ by substitution in eqn~(\ref{eq:profile-bg}) yields:
 \begin{eqnarray} \ln(\cC_w) + \frac{1}{\psi} \ln
\left(\frac{1 + \psi(1 + \alpha {\rm e}^{-w/d\s{h}}\cC_w)}{1 + \psi(1 + \alpha)}
\right) = - w/d\s{t}. \label{eq:profile-bg-Casper}
\end{eqnarray} 
We plot in Fig.~\ref{fig:Casper} the corrected exit intensity $\cC_w$ implied by (\ref{eq:profile-bg-Casper}).  For very intense beams, $\alpha \gg 1$, the corrected intensity tends to 1.  The dye is bleached and absorption is predominantly by the host.  The true profile is then Beer-like and the correction is accurate.
At lower intensities the correction is also seen to work surprisingly well.  The upper of each curve for a given reduced  thickness $w/d\s{t}$ is for no host absorption, representing the pure system being examined.  The corrected curve (lower in each case) is close to the ideal curve, even for the very heavy absorption $w/d\s{h} = 0.3$ in the illustration.  For smaller absorptions the corrected and the ideal curves are practically indistinguishable.  Thus the straightforward correction method should give a good estimate of the underlying non-linear absorptive processes.

Much of the discussion has been of intensities with reference to the
ideal Beer intensity, for instance Fig~\ref{fig:predict}(a).  Beer
behavior obtains when intensities are low enough that the \trans
population is little reduced.  This also obtains at short times when
as yet little conversion has taken place.  An output power $\cI_w =
\exp(-w/d\s{t})$ must emerge and has the value of the vertical
axis intercept in Fig~\ref{fig:predict}(a) (and is also the same as the reduced weak-beam emergent power). Thus, for a given power
$\alpha$ and thickness $w/d\s{t}$, the initial emergent power must
rise to the higher value characteristic of the curve in question.
The dynamics by which a spatial bleaching pattern is established is
non-linear and complicated, see comments below eqn~(\ref{eq:atten}).

One can also propose that bleaching can lead to an emergent power
that \textit{diminishes} in time from the initial, Beer value if the
\cis species have a higher absorption than the \trans.
\begin{figure}[!t]
\centering\resizebox{0.48\textwidth}{!}{\includegraphics{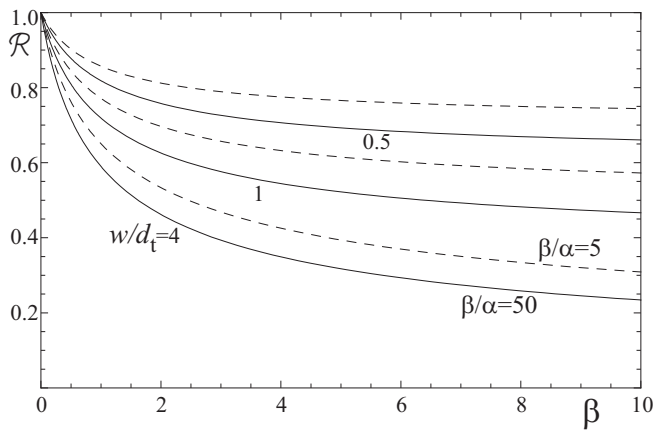}}
\caption[]{Intensity at the back surface scaled by the Beer intensity expected at the back face, $\cR=\cI_{w}/\cI\s{B}$ as a
function of the incident intensity scaled by the intensity characteristic of the \cis state, that is $\beta = I_0/I\s{c}$ for
cantilever thicknesses $w/d\s{t}= 4$, 1 and 0.5 with  $\beta/\alpha = 50$ (solid) and $\beta/\alpha = 5$ (dashed) pairs of lines for each thickness. The emergent intensity is lower than the Beer value, in contrast to the rest of the paper where bleaching enhances penetration.} \label{fig:anti}
\end{figure} A higher \cis absorption arises if the incident light
frequency is shifted to be closer to the \cis absorption line.  One
can rearrange  eqn~(\ref{eq:intensity}) by evaluating it at $x=w$
and noting that $-w/d\s{t} = \ln(\cI\s{B})$, where $\s{B}$ denotes
the Beer, $t=0$ reduced intensity.  Thus denoting by $\cR$ the ratio
$\cI_w/\cI\s{B}$, one has: \bea
\ln(\cR)&=& \frac{1- \alpha/(\eta\beta)}{(1+\eta)/\eta}\ln\left(\frac{1+\beta' \cI_{B}R}{1+\beta'}\right).
\label{eq:antibleach} \eea It is evident that the $\ln$ on the right
hand side is always negative since $\cI_{B}R = \cI_w \le 1$ and
hence if $\alpha < \eta\beta$ then the left hand side is also
negative, that is $\cR <1$ -- the intensity drops below the Beer
value rather than rising with time and increasing power. The
criterion, referring back to the ratio of the quantum efficiencies,
amounts to $d\s{c} < d\s{t}$, that is, attenuation due to \cis is
greater than that due to \trans.   Sample thickness is encoded in
$\cI_{B} = \exp(-w/d\s{t})$. Solutions to eqn~(\ref{eq:antibleach})
are shown in Fig~\ref{fig:anti}. Care is needed in interpreting this
figure: as thickness increases, the absolute amount of light passing
through the sample is greatly reduced since the Beer normalisation
to $\cI_w$ in $\cR$ is becoming exponentially small. This reverse
effect of photo-bleaching maybe useful in occasions where at high
intensities it  is desirable to have a smaller fraction of the light
transmitted (but as noted above, the absolute amount of light
transmitted is still increasing with incident power). It is possible
that biological systems depend on this behavior.

\vspace{.4cm} Acknowledgement.  We are grateful for discussions with
David Statman, and to the EPSRC for funding.


\end{document}